# Commentary about the paper " Hypertractions and hyperstresses convey the same mechanical information Continuum Mech. Thermodyn. (2010) 22:163–176 " by Prof. Podio Guidugli and Prof. Vianello and some related papers on higher gradient theories

**Francesco dell'Isola[1], Pierre Seppecher[2]**

[1] Università di Roma La Sapienza
Dipartimento di Ingegneria Strutturale e Geotecnica
Via Eudossiana 18 Roma Italia
e-mail: `francesco.dellisola@uniroma1.it`
Laboratorio Strutture e Materiali Intelligenti Fondazione Tullio Levi-Civita Via S. Pasquale snc Cisterna di Latina Italia

[2] Institut de Mathématiques de Toulon U.F.R. des Sciences et Techniques, Université de Toulon et du Var, Avenue de l'Université, BP 132, 83957 La Garde Cedex e-mail: `seppecher@imath.fr`

The date of receipt and acceptance will be inserted by the editor

*In memory of Prof. Carlo Gavarini, who recently passed away. He assured a visiting grant to Rome for P.S. in 1996, during which the paper [15] was written.*

**Abstract** In this commentary we try to make clearer the state of the art concerning the relation between mechanical contact interactions and the different notions of stresses. We emphasize the importance of the concept of virtual displacements. Its role has been recognized in Mechanics and in Continuum Mechanics long ago (see e.g. [59], [45], or [12], [13]) and it is central as well when starting with an expression of the power expended by internal stresses and deducing the form of contact interactions as when starting with some form of the contact interactions and developing a representation theorem for these contact interactions based on the Cauchy tetrahedron construction.

*Send offprint requests to*: francesco.dellisola@uniroma1.it
*Correspondence to*: francesco.dellisola@uniroma1.it

**Key words**    contact interactions, Cauchy tetrahedron, internal stresses

# 1 Short review of some relevant papers on the subject

In the last fifty years it has been widely recognized that in order to describe a wealth of physical phenomena it is needed to introduce mechanical theories which take into account contact actions more complex than those considered in continuum mechanics after Cauchy. Some well-known contributions in this regard are given in the papers [56], [37], [26], [55], [54].

More recently it has been recognized that second or even higher gradient models are needed when continuum models are introduced for describing systems in which strong inhomogeneities of physical properties are present at eventually different lenght scales (see e.g. [2],[44],[40]).

Actually, immediately after the development of the Cauchy format of continuum mechanics, a first relevant generalization in the aforementioned direction was conceived by Eugène and François Cosserat, but their efforts were not continued until late in XX century. Cosserat described continuum bodies in which contact actions were to be modelled not only by means of surface forces, but also by means of surface couples. The conceptual differences between Cauchy-type continuum mechanics and Cosserat-type continuum mechanics were relevant, and the second one could not be obtained by means of simple modifications of the first one. The remarkable mathematical difficulties confronted by Cosserat rendered their work difficult to be accepted, and for a long period their results were nearly completely ignored. This circumstance can be easily understood: the structure of Cosserat contact actions is complex. Indeed in Cosserat continua one needs, together with Cauchy stress tensor also a Couple stress tensor, for representing contact Couples. Moreover Cosserat used systematically the Principle of Virtual Work in a form which, although they recognized to go back at least to Lagrage, has been only recently recovered see e.g.[25], [41]. It is an interesting topic by itself to discuss how many times the Principle of Virtual Works and various Minimization Principles have been rediscovered and reformulated: we limit ourselves to cite [45] or [59] for further details.

# 2 A first method for extending Cauchy model for continuous bodies

In order to develop continuum mechanics going beyond the Cauchy format it is possible to use at least two different approaches.

The most simple of them, used also by Cosserat, starts by postulating how the power expended by internal actions in a body depends on the "virtual" velocity field and its gradients. Starting from this postulate one can deduce, by means of a successive application of the theorem of divergence, i.e. by means of several iterative integrations by parts, what are the contact actions which can be exerted at the boundary of the considered body.

Hence, this method starts from the notion of stress tensors and deduces from it the concept of contact actions.

It is based on the D'Alembert Principle of Virtual Work and has been resumed by Paul Casal ([?]) and subsequently by Paul Germain, in his enlightening papers ([26],[27]). This Principle is undoubtedly a great tool in Mechanics which has not been improved since its original first and "standard" formulation. It is not clear in which sense [25] use a "non-standard" version of the Principle of Virtual Work (or Virtual Powers). Indeed already Cosserat assumed that the internal power expended on virtual velocities equals the external power (including inertial terms) for every virtual velocity fields and for every subbody (suitably regular) of considered deformable body.

This is generally a position generally mantained in the literature. For instance in [3], adopting the same spirit as in [14] and in [15], it is stated that:

*In particular, the approach by means of the theory of distributions, mentioned by Germain himself but not fully developed, is here adopted from the beginning. Clearly, in order to obtain deeper results such as the Cauchy Stress Theorem, some extra regularity has to be assumed. Note that a power depends in general from two variables, the velocity field and the subbody. So it is a bit more complex than a mere distribution.*

Indeed in [14] and in [15] the starting assumptions concerning contact actions postulates that for every subbody of considered body the powers they expend on a generic velocity field is a distribution (i.e. a linear and continuous functional on velocity fields). Subsequently in [14] and in [15] it is postulated the quasi-balance of power (as formulated by [39]) and by using different polynomial test velocity fields and different families of subbodies, the Cauchy construction for stress tensors is obtained.

The works of Germain have been taken up again and again, (e.g. in [25],[43],[41]) often rephrasing them without introducing any notable amelioration.

Germain, following a tradition set in France by André Lichnerowicz, uses the original version (and more efficient) absolute notation due to Levi-Civita. This version, at least in this context, is the most adapted, as many objects of different tensorial order are to be simultaneously handled. Sometimes those who are refraining from using the most sophisticated version of Levi-Civita absolute Calculus are lead to refer to the needed stress tensors and the related contact actions indistinctly using the names "hyperstresses" and "hypertractions". On the contrary Germain tries to convey through the nomenclature chosen the physical meaning to be attached to the new mathematical objects he is introducing: for instance he calls "double forces" the actions which are expending powers on the velocity gradient in the directions which are normal to the surfaces of Cauchy cuts. Germain then decomposes these "double forces" into "couples" and "symmetric double forces" recognizing that couples were already introduced by Cosserat. Germain's notation supports the mechanical and physical intuition contrarily to what does a generic nomenclature based on some "hyper" prefixes.

## 3 A second method for extending Cauchy model and its relationship with the first

The second method starts by postulating the type of contact action which can be exerted on the boundary of every "regular" part of a body and then proceeds by proving a "representation" theorem for the considered class of contact actions: the existence of stress tensors is deduced from the postulated form of contact actions and a "balance-type" postulate, based on physical grounds. In other words: to the "constitutive" assumption chosen for characterizing the class of contact action under consideration one must add a Principle of Balance: the contact actions have to be balanced by a bulk action. This is the method followed by Cauchy which is often considered as the foundation of Continuum Mechanics. The important contribution due to [39] is to have introduced the assumption of "quasi-balance" for powers, which generalized, in the most suitable way, the Euler-Cauchy Postulate used in Cauchy continuum mechanics.

*3.1 The mathematical difficulties presented by this second method*

As remarked explicitly in [15],[14],[3],[33] the mathematical difficulty to be confronted in order to establish a firm foundation for this second method relies on the dual dependence of power functional on velocity fields and on subbodies of the considered continuum. It is obvious, starting from physical plausibility considerations, that power functionals must be regarded as distributions on the set of test functions represented by the admissible velocity fields (see e.g. [26], [15], [3]).

A foundamental results due to Schwartz allows for representing distributions (with compact support) as finite sums of derivatives of measures [46]. When (as it is important for considering contact actions) the distribution is concentrated on a smooth submanifold of threedimensional Euclidean space, then the derivatives to be considered are only those "normal" or "transversal" to the submanifold itself. Unfortunately in Schwartz it is not considered a representation theorem for families of distributions "attached to" the family of measurable subset of a given measurable set.

The efforts of [3], [21], [33], [21] are directed, with remarkable results, to the search of such a generalized Schwartz representation theorem.

Also of relevance is the problem arising when one must define generalized "stresses" having a flux which allows the representation of contact action and a divergence to be used for formulating bulk "local" form of balance laws.

This problem has been also addressed with some interesting results (see [32], [22], [53], [53])

*3.2 The two methods can be reconciled.*

During a long period the first method has been rejected by many researchers and it is lucky for advancement of science that its power has been, in the last decade, finally nearly unanimously accepted.

Moreover the two methods can be reconciled. Indeed the equivalence of the two methods has been explicitly established by Cauchy him-self and precised by Noll, for First Gradient Theories.

The same equivalence has been proven for the so called Second Gradient Theories, i.e. for theories in which the internal power is a second order distribution: this results has been obtained in the sequence of papers [39],[14],[15]. The relationship between the concept of contact line force and surface double forces was there mathematically proven obtaining also a representation formula relating the two concepts.

*3.3 The firm foundations of second gradient theories*

During the last century, second gradient theories have been the subject of controversies. The works of E. and F. Cosserat have been underestimated and misunderstood because these controversies were not solved in a sufficiently clear way until the papers of Paul Germain.

These controversies were due to two facts : the first is that second gradient theories are not compatible with the concept of stress as formalized by Cauchy. The second is the strong belief of a large part of the mechanics community that Cauchy stress was the only and universal framework in which all theories of continuous material should take place. To overcome the diffuculties different more or less exotic concepts were invoked. Let us quote the notion of interstitial working [23], the theory of extended thermodynamics [38], the notion of configurational forces [29], the so-called non-standard principle of virtual work [25].

Now time has come when the solid foundation of second and higher gradient theories must be recognized. Unfortunately it seems that the fundamental connection between the two aforementioned methods (and the available proof of the existence of this connection at least for those materials which were called by Germain *second gradient materials*) seems still not well understood in part of the mechanics community, while it has been considered as established by others (see e.g. [35], [36], [24]).

– The "*full set of representation formulae not only, as is relatively easy, for tractions and hypertractions in terms of stresses and hyperstresses, but also, conversely, for stresses and hyperstresses in terms of diffused and concentrated tractions and hypertractions ... generalizing the corresponding formulae for simple materials*" invoked in [43] has been established by Paul Germain ([26],[28]) in the seventies (extending the results of Casal [9], [10]) and has sufficiently and with more or less succes been rephrased or exploited (see e.g. [31],[50], [8], [51],[17],[18], [25]) and even recalled in the textbook [24]. Applications of second gradient theories to the mechanics of porous media is proposed for instance (but many other references could be given) in [11], [48] [19] where many results listed as novel in [43] are exploited.

- The fact that "*since we work in a nonvariational setting, our results apply whatever the material response*" is already contained in the cited original papers of Paul Germain. So is the fact that "*without edge tractions, both internal and external it would not be possible to arrive at the complete representation formula for the hyperstress in terms of hypertractions* " or the "*interesting feature of second-gradient materials is that, if bodies and subbodies having non everywhere smooth boundary are considered, then edge forces, i.e., line distributions of hypertractions are to be expected* ".
- The characterization of the class of second gradient materials for which contact edge action are always vanishing was already shown and exploited before the paper [43], contrarily to what is there stated: *we provide a new proof of the following not very well-known fact in the theory of second-gradient materials: if edge tractions are constitutively presumed null on whatever edge, then the hyperstress takes a very special form whose information content is carried by a vector field. We surmise that inability to develop edge interactions be characteristic of certain second-gradient fluids, an issue that we take up in a forthcoming article (..), continuing a line of thought proposed by Podio-Guidugli.* Indeed, this results, rather obvious, is obtained in exactly the same way in [15], Remark 3, pag. 48 and systematically exploited in the applications of second gradient theory presented in [19], [48]. Some interesting consideration about this point are already available in [51] togheter with some consideration about third gradient fluids. Remark that Equation (35) on pag. 173 in [43] for instance is exactly equal to Equation (18) pag. 6612 in [47] or to Equation (13) pag.107 in [49].
- Contrarily to the claim that "to the knowledge" of the authors of [43] "*a rigorous interaction theory accommodating such a nonstandard behavior remains to be constructed; interesting attempts in this direction have been carried out by Dell' Isola and Seppecher*", we are not aware of any precise criticism of the rigorous interaction theory which has been developed in [14] and [15]. For sake of simplicity we have only considered in these papers contact interactions the type of which limits the theory to a second gradient one. We show in the forthcoming work [16] how easy it is to extend the study to any type of distribution interactions and get thus any higher order gradient theory.
- It is not clear if the authors [43] are really aware of the assumptions and theorems presented in [39], [14] and in [15] which supply, in our opinion, the demanded Cauchy-like construction for second gradient materials. Indeed in [43] one reads "*Relations (7) and (8) are also arrived at when, as is customary, only tractions on body parts are introduced, because stress is constructed **à la Cauchy** as a consequence of balance* *of tetrahedron-shaped parts. The Cauchy construction is the pillar on top of which the standard theory of diffuse (i.e., absolutely continuous with respect to the area measure) contact interactions stands. For complex (i.e., nonsimple) material bodies, a Cauchy-like construction has been attempted often, but not achieved so far, to our knowledge.* On the contrary in the Conclusions of the paper [15] one can read: *The most important concepts introduced in this paper are: (i) the concept of quasi-balanced power of contact force distribution and (ii) that of prescribed shapes. They allowed us to develop a system of axioms "**à la Cauchy**' for continua in which edge contact forces are present.*

The connection between internal power and the power expended by external actions has not been yet completely established for a generic Nth Gradient Theory, although interesting and useful considerations can be found in [20], and also in [42].

In [16] and in a forthcoming paper it will be shown how the work started in ([39], [14], [15]) can be continued. The aim in these lecture notes will be to give a firm framework to those researchers which need to deal with more complex contact actions (for instance "wedge forces"), wish to refrain from using the Principle of Virtual Power and instead prefer to adopt an approch based on "contact interactions" rather than on "virtual power expended on virtual velocity fields". Indeed the original ideas presented in ([15]) can be rather easily extended in order to treat the case of all types of contact distributions: more precisely the Cauchy tetrahedron argument can be generalized to prove that all types of mechanical contact actions can be represented in terms of generalized stress tensors.